\documentstyle[12pt,psfig]{article}

\begin{document}
\newcommand{\be}{\begin{equation}}
\newcommand{\ee}{\end{equation}}
\newcommand{\bea}{\begin{eqnarray}}
\newcommand{\eea}{\end{eqnarray}}
\renewcommand{\topfraction}{1.}

\title{Stabilization of 2D Quantum Gravity by
branching interactions}

\author{{\bf Oscar Diego}\thanks{e-mail: imtod67@cc.csic.es} \\
        {\em Instituto de Estructura de la Materia, CSIC } \\
        {\em Serrano 123, 28006 Madrid} \\
        {\em Spain } }

\date{\mbox{ }}

\maketitle

\thispagestyle{empty}

\begin{abstract}

In this paper the stabilization of 2D quantum gravity by
branching interactions is considered. The perturbative
expansion and the first nonperturbative term of the
stabilized model are the same than the
unbounded matrix model which define pure
gravity, but it has new
nonperturbative effects that survives in the
continuum limit.

\end{abstract}

\begin{flushright}
\vspace{-13.5 cm} {IEM-FT-95/112}
\end{flushright}
\baselineskip=21pt
\vfill
\newpage
\setcounter{page}{1}

{\bf 1.} The topological expansions of 2D quantum gravity coupled to unitary
conformal matter with $c \le 1$, are non Borel summable\cite{BOREL}, and
therefore, is not possible extract from them
nonperturbative definitions of 2D quantum gravity.
But, if one found some quantum field theory
with the same perturbative expansion than the topological
expansion and with sensible instantonic configurations
related to the non Borel summability,
the quantum field theory would be the nonperturbative definition
of 2D quantum gravity.

The perturbative series of matrix models are
the same than the topological expansions of 2D quantum
gravity\cite{MATRIX},
but unfortunately they are ill defined nonperturbatively
because their
potentials are unbounded from below. There are several
well defined models with the same perturbative expansions than the
matrix models, unfortunately not all of them verify
the minimal conditions of every sensible quantum field theory:

a) Reality of the physical observables; for instance,
analytical continuations gives complex observables\cite{COMPLEX}.

b) Instantonic configurations related to the
asymptotic behaviour of the topological expansion: the
origin of nonperturbative behaviours unrelated to the perturbative
expansions are arbitrary restrictions on the configuration space,
for instance, in unitary matrix models the eigenvalues are
restricted to the real positive axis\cite{DALLEY},
and in some nonperturbative definitions of $c=1$ matrix model the
eigenvalues are restricted by an infinite wall. And this restrictions
have not physical meaning.

For gravity without matter
only the stochastic stabilization
verify the above two conditions\cite{STAB},
but unfortunately is very difficult to extend
it to the case of 2D quantum gravity coupled to $c = 1$ matter.

In this paper a new stabilization method for 2D quantum gravity
without matter is proposed,
which verify conditions a) and b).

The new stabilized models are constructed by adding branching
interactions to the unbounded potentials, for example:
\be
V = Tr ( \Phi^{2} - g \Phi^{3}) + \frac{\alpha}{N} Tr \left (\Phi^{2} \right
)^{2}.
\label{eq:pra}
\ee
For $N$ fixed and finite, $V$ is bounded from below if $\alpha$ is positive,
but in
the planar limit the Hartree approximation is exact and the
potential becomes:
\be
V_{Hartree} = Tr ( \Phi^{2} - g \Phi^{3}) + \alpha \omega Tr (\Phi^{2})
\ee
where $\omega$ is the vacuum expectation value of $\frac{1}{N}
Tr \Phi^{2}$, therefore
in the planar limit the effective potential is the same that the
potential of the unbounded matrix model with a self consistent parameter,
and this model belongs to the universality class of pure gravity. Of
course matrix models with branching interactions
can have several critical points\cite{DAS,KLE,SUG,BRA},
but in each phase there are
only one vacuum. Therefore, in some region of the $(g,\alpha)$ phase space,
(\ref{eq:pra}) defines the topological expansion of 2D quantum
gravity.

This behaviour is very different from stabilization with trace terms,
for example:
\be
V = Tr ( \Phi^{2} - g \Phi^{3} + \alpha \Phi^{4} )
\ee
where $V$, for $\alpha$ positive, is also bounded from below. But in the planar
limit the potential is bounded from below, and in the pure
gravity phase there are almost two vacuum\cite{ARCS}, one is the vacuum of pure
gravity and the other has not sensible physical interpretation. In
fact, the well defined vacuum is probably the new vacuum, and the pure
gravity vacuum probably only exist in the perturbative expansion.

In\cite{KLE,SUG} has been claimed that, in the pure gravity
phase, the matrix models with branching interactions are equivalents
to the unbounded matrix models nonperturbatively. The argument is the
following:
first,  the partition function (\ref{eq:pra}) is rewritten:
\bea
Z & = & \int d \Phi \int d x e^{ \frac{ N^{2} x^{2}}{\alpha} }
\exp \{ - N Tr V_{x} \} \nonumber \\
V_{x} & = & V + 2 x Tr \Phi^{2}
\label{eq:prax}
\eea
then, the integration over $\Phi$ is performed and after the
double scaling limit the partition function becomes:
\be
Z = \int_{-\infty}^{\infty} d x \delta (x) Z_{0} (N^{4/5} ( g_{c} (x) -  g ) )
\ee
where $Z_{0}$ is the partition function of the unbounded matrix model.
Therefore in the double scaling limit both matrix models are equivalents.
But this argument is wrong because the integration first over $\Phi$ and then
over $x$ in (\ref{eq:prax})
is ill defined for finite $N$, while (\ref{eq:pra}) is well defined,
therefore the integrations $dx$ and $d \Phi$ do not commute for
$N$ finite.

I will show also that in the double scaling limit both integrations
$d \Phi$ and $d x$ do not commute. There will be new nonperturbative
contributions to the partition function when the integrations
over $\Phi$ and $x$ are performed
before the double scaling limit. This new contribution survives
in the double scaling limit.

Even the new matrix models verify both conditions a) and b),
they do not verify the Ward identities of matrix models
(loops equations, string equation, KdV flows).
In fact, it is not possible to found a stabilization that verify the
reality condition a), and the loop equations, string equation and all
the mathematical structures which arise in a perturbative analysis
of the matrix models\cite{DAVID}. In fact, I will show that the
new nonperturbative effects are related with the break down
of the symmetries of the matrix model.

{\bf 2.} Let be the matrix model:
\bea
Z & = & \int d \Phi \exp \{- N V \} \nonumber \\
V & = & Tr \{ \Phi^{2} - 2 \frac{g}{3} \Phi ^{3} \}
\label {eq:m1}
\eea
it is the simplest model which define 2D quantum gravity without matter
and is unbounded from below.

Perturbatively:
\be
\int d \Phi \sum_{a} \frac{\partial}
{\partial \Phi_{aa}} \{ exp (-N V ) \} = 0
\ee
therefore
\bea
\langle P(\Phi) \rangle & = & 0 \nonumber \\
P(\Phi) & = & Tr [ g \Phi^{2} - \Phi ]
\label {eq:e1}
\eea
this is the first equation of a set of identities
which are equivalents to the loops equations and Virasoro constraints.

Let us define the new matrix model
\bea
Z_{\alpha} & = & \int d \Phi \exp \{ - N W_{\alpha} \} \nonumber \\
W_{\alpha} & = & V (\lambda ) + \frac{\alpha}{N} \left [ P(\Phi) \right ]^{2}
\label {eq:stb}
\eea
where $\alpha$ is positive.

Integrating over the angular variables the new matrix model
becomes:
\bea
Z_{\alpha} & = & \int \prod_{i=1}^{N} d \lambda_{i}
\exp \{ - N S_{eff} \} \nonumber \\
S_{eff} & = & \sum_{i=1}^{N} V (\lambda_{i}) +
\frac{1}{N} \sum_{i \neq j} \ln | \lambda_{i} - \lambda_{j} |
+ \frac{\alpha}{N} \left [ \sum_{i=1}^{N} P(\lambda_{i}) \right ]^{2}.
\eea

For finite $N$, $S_{eff}$ is bounded from below: when $\lambda_{i}$
goes to infinity, the effective action increases as $\alpha g \lambda_{i}^{4}$.
Hence, $Z_{\alpha}$ is well defined.

In the planar limit, $N \rightarrow \infty$, the branching interaction is
replaced by the Hartree term
\be
\frac{1}{N} \left ( Tr P \right )^{2} \rightarrow \langle P \rangle Tr P
\ee
and the eigenvalue density $\rho$ in the large $N$ limit
is given by the saddle point condition:
\bea
- g \lambda^2 + \lambda \left ( 1 + 2 \alpha g \langle P \rangle  \right )
- \alpha \langle P \rangle +
\int d \mu \frac{ \rho (\mu) }{\lambda - \mu} & = & 0
\nonumber \\
\int d \mu \rho(\lambda) \left ( g \mu^2 - \mu \right ) & = &
\langle P \rangle  \nonumber \\
\int d \mu \rho ( \mu ) & = & 1
\label{eq:point}
\eea
and from (\ref{eq:e1}) there is one
solution with $\langle P \rangle = 0$,
and with the same
planar eigenvalue density
than the unbounded matrix model (\ref{eq:m1}).
This solution is unique because the above equations are
the saddle-point equations for an unbounded
matrix potential with only one local minimum.
Therefore, in the planar limit, the stabilized matrix
model is independent on $\alpha$, and the
continuum limit is defined by a
critical coupling constant $g_{c}$ independent on $\alpha$.

Following reference\cite{KLE} the partition function $Z_{\alpha}$
can be rewritten as:
\bea
Z_{\alpha} & = & \int d \Phi \int d x
exp \{ - N {\bar{S}}_{eff} \} \nonumber \\
{\bar{S}}_{eff} & = & \sum_{i=1}^{N} V (\lambda_{i}) +
\frac{1}{N} \sum_{i \neq j} \ln | \lambda_{i} - \lambda_{j} |
+ 2 x \sum_{i=1}^{N} P(\lambda_{i}) - \frac{N x^2}{\alpha}
\label{eq:kpo}
\eea
because all terms in the effective action ${\bar{S}}_{eff}$ are
order $N$, the saddle point condition is given by:
\bea
\frac{\partial {\bar{S}}_{eff} }{\partial \lambda_{k} } & = & 0 \nonumber \\
\frac{\partial {\bar{S}}_{eff} }{\partial x } & = & 0
\eea
and is trivial to check that the above equations are the
same than (\ref{eq:point}).
The saddle point value for $x$ is
\be
x_{s} = \alpha \langle P \rangle.
\ee

At subleading order in the expansion $1/N$,
$Z_{\alpha}$ must depends on $\alpha$, but in
the double scaling limit only survives the dependence on
$\alpha$
through the value of the critical coupling constant $g_{c}$,
which is independent
on $\alpha$. Hence the new matrix model $Z_{\alpha}$, has the same
topological expansion than the unbounded matrix model (\ref{eq:m1}).

Equation (\ref{eq:e1}) is perturbative, nonperturbatively $\langle P
\rangle$
must be different from zero. This means that the symmetries of the
matrix model
must break down by nonperturbative corrections.

$\langle P \rangle$ can be calculated as follows:
\be
\langle P \rangle = \int d \Phi P (\Phi) \exp (-N W_{\alpha})
\label {eq:alp}
\ee
the first nonperturbative contribution to $\langle P \rangle$
is given by the instantonic configuration: $N-1$
eigenvalues are placed in the support of $\rho$ and one
eigenvalue $\lambda_{T}$ is
placed on the top of the effective potential for one eigenvalue:
\be
\Gamma_{\alpha} =  -N \{ \lambda^{2} - 2 \frac{g}{3} \lambda^{3} + 2 \alpha
\langle P \rangle ( g \lambda^{2} - \lambda )
+ 2 \int d \mu \rho (\mu) \ln | \lambda
- \mu | \}
\ee
and only the isolated eigenvalue $\lambda_{T}$ gives the first nonperturbative
contribution to (\ref{eq:alp})
\be
\langle P \rangle = \frac{1}{N} ( g \lambda_{T}^{2} - \lambda_{T} )
\exp (-N \Gamma_{\alpha}(\lambda_{T})).
\ee
The instantonic configuration $\lambda_{T}$ is then
given by the condition
\be
\frac{d}{d \lambda} \left ( \Gamma_{\alpha} \right )_{\lambda = \lambda_{T}}= 0
\ee
and at leading order the above equation becomes:
\be
\lambda - g \lambda^{2} + \int d \mu \frac{ \rho (\mu) }{\lambda - \mu} = 0
\ee
which is exactly the same equation for the instantonic configuration of
the unbounded matrix model. Therefore, in the double scaling limit,
the first nonperturbative correction is the same in both matrix
models:
\be
\delta Z_{\alpha}^{0} = \exp \left ( -N \Gamma_{\alpha = 0 }
(\lambda_{T} ) \right )
\ee

{}From (\ref{eq:alp}),  the instantonic configuration with one eigenvalue
placed
at the top of the effective potential gives the leading
term depending on $\alpha$
that survives in the double scaling limit:
\be
\delta Z_{\alpha}^{1} = \exp \left ( - 2 \alpha ( g \lambda_{T}^{2} -
\lambda_{T} )
\exp (- N \Gamma_{\alpha=0}(\lambda_{T}) ) \right ).
\label{eq:cor}
\ee
this factor is not universal but is finite in the double scaling limit and
depends on the renormalized coupling constant. Therefore the partition
function in the double scaling limit is given by:
\be
Z_{\alpha} = Z_{0} + \delta Z_{\alpha}^{0} \delta Z_{\alpha}^{1}
\ee
where $Z_{0}$ is the perturbative part of the partition
function and is independent on $\alpha$.

The same results can be obtained from
the integral representation (\ref{eq:kpo}).
But the integral over $x$ must be
performed at the same time that integration over the
eigenvalues, otherwise the new correction $\delta Z_{\alpha}^{1}$
cannot be calculated from (\ref{eq:kpo}).

This behaviour is different from stabilization with trace
terms, for instance:
\be
V = Tr(\Phi^2 - 2 \frac{g}{3} \Phi^3 + \frac{\alpha}{N} \Phi^4)
\label{eq:tra}
\ee
where the planar limit is also independent of $\alpha$ and therefore
the perturbative double scaling must be independent on $\alpha$.
But the nonperturbative dependence on $\alpha$ which survives in the
double scaling limit is
\be
\delta Z_{\alpha}^{1} = \exp \left ( \alpha \lambda_{T}^4 \right )
\ee
which is independent on the renormalized
coupling constant, this fact suggest that the scaling part
of the partition function that depends on the cosmological
constant does not depend on the stabilized term of the potential.
Therefore even the model (\ref{eq:tra})
is well defined for finite $N$, the double scaling
limit is probably ill defined nonperturbatively.

{\bf 3.} The conclusions that follows from this work are:

The matrix model with branching interactions (\ref{eq:stb})
defines the topological expansion and the first nonperturbative
correction of unbounded matrix model (\ref{eq:m1}).
As in the
stochastic stabilization, there are only one perturbative
vacua.

The branching term gives
a new nonperturbative correction to the partition function
that survives in the double scaling limit
but they are not universal. Therefore, there are an infinity number of
well defined models that verify the two conditions a) and b). In order
to fix only one model, additional assumptions are needed.

This new nonperturbative correction depends on the renormalized
cosmological constant, and this support the hypothesis that the
stabilized model is well defined in the double scaling limit because the
scaling part of the
partition function must depends on the stabilized term of the matrix
potential.

The integral representation (\ref{eq:kpo}) is ill defined and
the integration over $x$ and $\Phi$ commute only perturbatively.
Unfortunately the integral representation of\cite{KLE,SUG} can not
be used to perform more rigorous nonperturbative computation.

The new matrix model does not verify the usual loops equations
and from the identities
\be
\int d \Phi \frac{\partial}{\partial \Phi_{ab}} \left [ \left (
\exp(L\Phi) \right )_{ab}  \exp (-N W_{\alpha}) \right ] = 0
\ee
one can extract new loops equations:
\be
N V^{\prime} \left ( \frac{\partial}{\partial L} \right )
\langle W(L) \rangle = \int_{0}^{L} d J \langle W(J) W(L-J) \rangle
- 2 \alpha N P^{\prime} \left ( \frac{\partial}{\partial L} \right )
\langle P W(L) \rangle.
\ee
there is an open question the physical meaning of this new loops
equations.

Therefore the stabilization with branching interactions may be
a sensible definitions of 2D quantum gravity, and can
be extended to the $c=1$ case, whereas the stochastic stabilization
for $c=1$ matrix model is given by a non solvable model.

The results of this paper can be extended to arbitrary branching
interactions, for instance:
\be
W_{\alpha} = V + \frac{\alpha}{N} Tr \Phi^{2} Tr \Phi^{2}.
\ee
This matrix model has several phases depending on the
value of $\alpha$, and there is one phase belonging
to the universality class of pure gravity. Of course,
in this phase
the critical coupling constant depends on $\alpha$.

Even for the pure gravity phase
there is a new nonperturbative contribution to the
partition function, which is not
universal but is finite in the double scaling limit.
This new term is given by the nonperturbative
contribution to $\langle Tr \Phi^{2} \rangle$.

{\bf Acknowledgment}

I would like to thank University of Santiago for the
financial support and the Particle Physics
Department for the hospitality during part of this work.

\vfill

\pagebreak

\end{document}